\documentclass[aps,prl,twocolumn,longbibliography,epsfig,floats,preprintnumbers,floatfix]{revtex4-1}
\usepackage{graphicx}

\usepackage{amsmath}
\usepackage{amssymb}
\usepackage{bm}
\usepackage{color}
\usepackage{sidecap}

\usepackage[utf8]{inputenc}

\definecolor{NatureBlue}{rgb}{0.012,0.3,0.63}
\definecolor{Red}{rgb}{0.81,0.09,0.13}

\newcommand{\os}{\bm\sigma\!_s}
\newcommand{\osf}{\bm\sigma'\!\!_s}
\newcommand{\ot}{\bm\sigma\!_t}
\newcommand{\otf}{\bm\sigma'\!\!_t}
\newcommand{\oii}{\bm\sigma\!_i}
\newcommand{\oif}{\bm\sigma'\!\!_i}
\newcommand{\dir}{\rightarrow}
\newcommand{\dil}{\leftarrow}
\newcommand{\diu}{\uparrow}
\newcommand{\did}{\downarrow}

\begin{document}
\title{First Encounter Kinetics of Active Particles}
\author{Marc Robin Riebes}
\affiliation{Department of Theoretical Physics and Center for Biophysics, 
Saarland University, 66123 Saarbr\"ucken, Germany}
\author{Reza Shaebani}
\email{shaebani@lusi.uni-sb.de}
\affiliation{Department of Theoretical Physics and Center for Biophysics, 
Saarland University, 66123 Saarbr\"ucken, Germany}

\begin{abstract}
First-encounter processes between moving entities govern the timescales of interactions across 
a broad range of biological, physical, and synthetic systems. Despite extensive studies of 
diffusive first-passage processes, first encounters between independently moving active particles 
remain largely unexplored. We develop an analytical framework for calculating the mean first-encounter 
time (MFET) of two active particles in a bounded domain with independently tunable persistence 
and mobility asymmetries. By formulating the dynamics in relative coordinates while retaining 
the instantaneous directions of both particles, we overcome the difficulty of reducing two 
independently persistent walkers to a single effective active particle. We show that mobility 
fundamentally reshapes the optimal encounter strategy. Remarkably, depending on the individual 
persistence values of the two particles, mobilizing an initially immobile particle can reduce 
the MFET even below the globally optimal static value. However, this advantage is maximal at 
a persistence-dependent optimal mobility and vanishes as the particles approach equal mobility. 
Optimal encounters favor asymmetric persistence, with the faster particle adopting high and the 
slower particle low persistence, while increasing mobility symmetry reshapes the optimization 
landscape toward a competing swapped configuration. We further identify discontinuous transitions 
between competing mobility optima in the ballistic regime. The analytical predictions are in 
excellent agreement with numerical simulations, establishing a general framework for first-encounter 
kinetics in active matter and providing a basis for studying more complex encounters involving 
heterogeneous motility and multiple active agents.
\end{abstract}

\maketitle

\noindent{\bf Introduction} 
\smallskip\smallskip

\noindent First-encounter processes underlie interaction kinetics across a broad range of stochastic systems,
from diffusion-controlled chemical reactions and molecular recognition to biological interactions
and ecological dynamics \cite{Benichou14,Metzler14,Loverdo08,Normanno15,Benichou10}. Since the
seminal work of Smoluchowski \cite{Smoluchowski18}, encounter kinetics between diffusing particles
have been extensively studied within the framework of first-passage theory \cite{Redner01},
providing the theoretical foundation for reaction-rate theory and stochastic transport. In many
situations, the two-body encounter problem can be mapped onto the motion of a single particle
whose effective diffusion coefficient is the sum of the individual ones, explaining why target
mobility generally accelerates encounters \cite{Giuggioli13,Borrego09,Yuste08,Bray02,Bramson88,
Weng17,Holcman09}. Although confinement, topological heterogeneity, or interparticle interactions
can modify this picture \cite{LeVot20,Oshanin09,LeVot22,Kreten26,Tejedor11b,Jose18,Schwarzl16,
Shaebani18,Teryoshin26,Kreten25,Gabel12}, existing theory remains largely rooted in passive
diffusion. Such a mapping, however, generally breaks down when the interacting particles are
active because they retain their independent orientational memory \cite{Sadjadi21}. This leaves
a largely unexplored class of first-encounter problems in which distinct motility characteristics
of both encounter partners are preserved.

The emergence of active matter has fundamentally extended first-passage theory beyond thermal
diffusion. Active particles exhibit first-passage dynamics governed by persistence, noise, and
confinement, leading to search strategies that differ qualitatively from their passive counterparts
\cite{Tejedor12,Shaebani22,Chupeau15,Guerin16}. A central result is that confinement gives rise
to an optimal persistence that minimizes the mean first-passage time to a stationary target by
balancing local exploration against long-range transport \cite{Tejedor12}. While this picture
has been generalized to correlated search strategies, stochastic resetting, complex environments,
and non-Markovian processes \cite{Guerin16,Shaebani20,Condamin07}, it is based on the assumption
that one encounter partner is an immobile target. Yet many encounters in active matter involve
independently moving active agents, including microswimmers, motile cells, active colloids,
microrobots, and self-propelled drug carriers seeking specific binding targets \cite{Bechinger16,
Najafi18,Palacci13,Kreten26b,Harris12}. A fundamental open question is whether the optimal
principles established for active first-passage to a static target remain valid when both
encounter partners move actively.

Here, we study first-encounter kinetics between independently moving active particles confined
in finite domains. We address the central question of how the mobility and persistence of both
encounter partners jointly determine encounter kinetics and optimal strategies. To this end, we
develop a backward master-equation formalism to solve the first-encounter problem for the relative
dynamics of two active particles, providing a general analytical framework for a previously
intractable class of first-encounter problems in active systems. Our results uncover a rich,
non-monotonic optimization landscape emerging from the interplay between persistence and mobility
asymmetries, with predictions in excellent agreement with extensive Monte Carlo simulations.

\begin{figure*}[t]
\centering
\includegraphics[width=0.55\textwidth,clip]{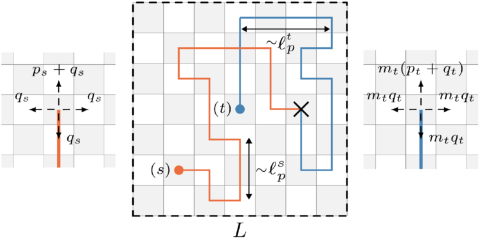}
\caption{Schematic illustration of the first-encounter trajectories of two independently moving
active particles $s$ (with mobility $1$ and persistence $p_s$; left inset) and $t$ (with
mobility $0{\leq}m_t{\leq}1$ and persistence $p_t$; right inset) on a square lattice.}
\label{Fig1}
\end{figure*}

\smallskip\smallskip\smallskip
\noindent{\bf Model} 
\smallskip\smallskip\smallskip

\noindent We consider two discrete-time persistent random walkers $s$ and $t$ on a $d$-dimensional periodic 
cubic lattice of volume $V{=}L^d$ (Fig.\,\ref{Fig1}). The formalism extends straightforwardly to 
arbitrary dimensions; for brevity, here we present the two-dimensional case ($d{=}2$). Particles 
are initially distributed uniformly in space with isotropic directions. Subsequently, at each 
time step, particle $i{\in}\{s,t\}$ retains its direction with probability $p_i$ or randomly 
selects a lattice direction with probability $q_i{=}\frac{1{-}p_i}{4}$. Thus, $p_i{=}0$ and $1$ 
correspond to diffusive and ballistic limits, respectively. The step length is set equal to 
the lattice spacing, which defines the unit of length. The persistence parameter is equivalently 
the directional correlation $p_i{=}\langle{\cos\theta_i}\rangle$, where $\theta_i$ denotes the 
turning angle between consecutive steps \cite{Sadjadi15,Burov13}. For unit lattice spacing, the 
corresponding persistence length $\ell_p^i$ satisfies $p_i\,{\propto}\,\mathrm{e}^{-1{/}\ell_p^i}$
\cite{Shaebani22b,Landau58}; for the present square lattice, $\ell_p^i{=}\frac{4}{3(1{-}p_i)}$.
To tune the mobility asymmetry, particle $t$ moves with the same step length as particle $s$ at
each time step but with probability $m_t{\in}[0,1]$ and remains stationary otherwise. Hence,
$m_t{=}0$ describes an immobile particle $t$, whereas $m_t{=}1$ corresponds to equal mobilities.
We choose $L$ odd to avoid even-odd sublattice decoupling and ensure encounters remain possible
for random initial positions at $m_t{=}1$.

\smallskip\smallskip\smallskip
\noindent{\bf First-encounter kinetics of two active particles} 
\smallskip\smallskip\smallskip

\noindent The independent persistence of the two particles precludes the conventional reduction to an effective
single walker. We overcome this obstacle by formulating the two-body dynamics exactly in relative
coordinates augmented by the instantaneous directions of both particles. At any time, the state
$u\,{=}\,(\bm{r},\os,\ot)$ is specified by the relative position $\bm{r}\,{=}\,(x,y){\in}Z\!_{_L}^{\,2}$
($Z\!_{_L}{=}\{0,1,{\cdot}{\cdot}{\cdot},L{-}1\}$), with coordinates taken modulo $L$, and instantaneous
directions $\os,\ot{\in}\{\dir,\dil,\diu,\did\}$. Encounters correspond to $\bm r\,{=}\,\bm0$, leaving
$16(L^2{-}1)$ nonabsorbing states $\mathcal{U}$. Denoting the step vector of particle $i$ with $\Delta(
\oii)$ and its conditional transition probability from direction $\oii$ to $\oif$ by $P(\oii|\oif)\,{=}
\,q_i\,{+}\,\delta_{\oii{,}\oif}p_i$, we introduce the mean first-encounter time (MFET) $\tau(\bm{r},\os,
\ot)$ for starting from the state $(\bm{r},\os,\ot)$. The evolution of $\tau$ then obeys the exact
backward master equation
\begin{equation}
\begin{split}
&\tau(\bm{r},\os,\ot)\,{=}\,(1{-}m_t)\sum_{\osf}P(\os|\osf)\tau\big(\bm{r} {+} \Delta(\os), \osf, \ot\big) \\
&{+}\,m_t\!\!\sum_{\osf}\!\sum_{\otf}\!P(\os|\osf)P(\ot|\otf)\tau\big(\bm{r} {+}\Delta(\os){-}\Delta(\ot),\osf,\otf\big) \\
&{+}\,1\,{-}\,\delta(\bm r)\,h(\os{,}\ot). \nonumber
\label{Eq:MasterEq}
\end{split}
\end{equation}
The last term on the right-hand side takes the mean return time for $\bm r\,{=}\,\bm0$ into account.
$h(\os{,}\ot)$ is determined self-consistently below. We next introduce the Fourier transform 
\[\widetilde{\tau}(\bm{k},\os,\ot)\,{=}\!\sum_{\bm{r} \in Z_{L}^{2}}\tau(\bm{r},\os,\ot)\text{e}^{-\text{i}\bm{k}{\cdot}\bm{r}},\]
with $k_{i}{=}\frac{2\pi n_{i}}{L}$ ($n_{i}\,{\in}\,\{0, 1, \ldots, L\,{-}\,1\}$). Introducing the $4{\times}4$
transition matrices $\bm{\underline{P}}_{s}$ and $\bm{\underline{P}}_{t}$ and the diagonal propagation matrices
$\bm{\underline{D}}_{s}(\bm{k})\,{=}\,\text{diag}(\text{e}^{\text{i}k_{x}},\text{e}^{-\text{i} k_{x}},\text{e}^{\text{i}
k_{y}},\text{e}^{-\text{i} k_{y}})$ and $\bm{\underline{D}}_{t}(\bm{k})\!\!=\bm{\underline{D}}_{s}(\bm{-k})$ for
the particles, the Fourier transform of the master equation yields
\begin{equation}
\begin{split}
\widetilde{\tau}(\bm{k},\os,&\ot)\,{=}\,V\delta(\bm{k})\,{-}\,h(\os{,}\ot)\,{+} \\
&\sum_{\osf}\!\sum_{\otf}\big[\bm{\underline{D}}_{s}(\bm{k})\bm{\underline{P}}_{s}\big]_{\os,\osf}\big[
\bm{\underline{R}}_{t}(\bm{k})\big]_{\ot,\otf}\widetilde{\tau}(\bm{k},\osf,\otf), \nonumber
\label{Eq:MasterEqFourier}
\end{split}
\end{equation}
with $\bm{\underline{R}}_{t}(\bm{k}){:=}(1{-}m_{t})\bm{\underline{I}}_{4}{+}m_{t}\bm{\underline{D}}_{t}
(\bm{k})\bm{\underline{P}}_{t}$.

\begin{figure*}
\centering
\includegraphics[width=0.88\textwidth]{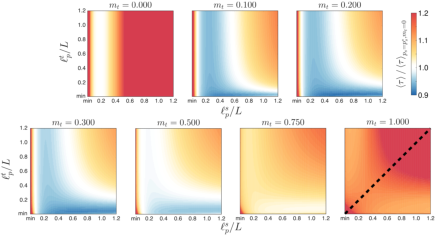}
\caption{Phase diagrams of the normalized mean first-encounter time (MFET) $\langle\tau\rangle$
in the ($\ell_p^s$,\,$\ell_p^t$) plane for $V{\simeq}10^4$ and different mobility parameters $m_t$. The MFET
is normalized by the globally optimal value for an immobile particle $t$ ($m_t\,{=}\,0$). Blue
(red) regions denote encounter times below (above) the optimal static benchmark. The sub-benchmark
blue region emerges at weak mobility and disappears as $m_t$ approaches unity. Increasing $m_t$
progressively reshapes the optimization landscape from a single asymmetric optimum toward two
competing optima related by interchange of the persistence lengths; at $m_t\,{=}\,1$, the
optimization landscape becomes fully symmetric about the dashed line.}
\label{Fig2}
\end{figure*}

To eliminate the explicit orientation indices, we collect the 16 Fourier components into the
lexicographically ordered vector $\widetilde{\bm\tau}(\bm{k})\,{=}\,\big(\widetilde{\tau}(\bm{k},\dir,\dir),
\widetilde{\tau}(\bm{k},\dir,\dil),\ldots,\widetilde{\tau}(\bm{k},\did,\did)\big)^{\top}$, and analogously
define $\bm{h}\,{=}\,(h(\dir,\dir),h(\dir,\dil),\ldots,h(\did,\did))^{\top}$. By constructing a set of 16 coupled
master equations for every $(\os,\ot)$ pair and introducing a $16{\times}16$ matrix $\bm{\underline{\Omega}}
(\bm{k})\,{=}\,\bm{\underline{I}}_{16}\,{-}\,\big[\bm{\underline{D}}_{s}(\bm{k})\bm{\underline{P}}_{s}\big]
\,{\otimes}\,\bm{\underline{R}}_{t}(\bm{k})$ (which is invertible for $p_s\,{<}\,1$ and $\bm{k}\,{\neq}\,
\bm{0}$), the set of master equations can be presented in the compact form $\widetilde{\bm\tau}(\bm{k})\,{=}\,
\,{-}\,\bm{\underline{\Omega}}^{-1}(\bm{k})\,\bm{h}$. The inversion of $\bm{\underline{\Omega}}(\bm{k})$
can be reduced to the diagonalization of two $4{\times}4$ matrices $\bm{\underline{D}}_{s}(\bm{k})\bm{
\underline{P}}_{s}$ and $\bm{\underline{D}}_{t}(\bm{k})\bm{\underline{P}}_{t}$, which results in
\[\bm{\underline{D}}_{s}(\bm{k})\bm{\underline{P}}_{s}\,{=}\,\bm{U}(\bm{k})\bm{\underline{\Lambda}}_{s}(\bm{k})
\bm{U}^{-1}(\bm{k})\] 
and 
\[\bm{\underline{D}}_{t}(\bm{k})\bm{\underline{P}}_{t}\,{=}\,\bm{V}(\bm{k})
\bm{\underline{\Lambda}}_{t}(\bm{k})\bm{V}^{-1}(\bm{k}).\] 
\noindent Here $\bm{U}$ and $\bm{V}$ are the respective
eigenvectors, $\bm{\underline{\Lambda}}_{s}\,{=}\,\text{diag}(\lambda_{1},\ldots,\lambda_{4})$,
$\bm{\underline{\Lambda}}_{t}\,{=}\,\text{diag}(\mu_{1},\ldots,\mu_{4})$, and $\lambda_{i}$ and
$\mu_{i}$ are the respective eigenvalues. Since $\bm{\underline{R}}_{t}(\bm{k})$ differs from
$\bm{\underline{D}}_{t}(\bm{k})\bm{\underline{P}}_{t}$ only by a linear combination with the
identity matrix, it has the same eigenvectors and can be written as $\bm{\underline{R}}_{t}(\bm{k})
{=}\bm{V}(\bm{k})\bm{\underline{\Lambda}}_{R}(\bm{k})\bm{V}^{-1}(\bm{k})$ with $\bm{\underline{
\Lambda}}_{R}(\bm{k})\,{=}\,\text{diag}\Big(1\,{-}\,m_{t}+m_{t}\mu_{1},\ldots,1\,{-}\,m_{t}\,{+}\,
m_{t}\mu_{4}\Big)$. Using the mixed-product property of Kronecker products then yields
\begin{equation}
\begin{split}
&\bm{\underline{\Omega}}^{-1}(\bm{k})\,{=} \\
&\left(\bm{U}{\otimes}\bm{V}\right)\text{diag}\Big(\frac{1}{1\,{-}\,\lambda_{i}\left(1\,{-}\,m_{t}\,{+}
\,m_{t}\mu_{j}\right)}\Big)_{i,j}\left(\bm{U}^{-1}{\otimes}\bm{V}^{-1}\right). \nonumber
\label{Eq:OmegaInverted}
\end{split}
\end{equation}

It remains to determine the zero Fourier mode. Introducing $\underline{\bm{G}}\,{=}\!\!\sum\limits_{
\bm{k}\,{\neq}\,\bm{0}}\bm{\underline{\Omega}}^{-1}(\bm{k})$ (where the sum runs over all nonzero
Fourier modes), the absorbing condition implies $\widetilde{\bm\tau}(\bm{0})\,{=}\,\underline{\bm{G}}
\,\bm{h}$. At $\bm{k}\,{=}\,\bm{0}$, we have $\bm{\underline{D}}_{s}(\bm{0})\,{=}\,\bm{\underline{D}}_{t}(\bm{0})\,
{=}\,\bm{\underline{I}}_{4}$ and hence $\bm{\underline{R}}_{t}(\bm{0})\,{=}\,(1\,{-}\,m_{t})\bm{
\underline{I}}_{4}\,{+}\,m_{t}\bm{\underline{P}}_{t}$. Substituting into the Fourier-transformed
master equation gives a self-consistent relation for $\bm{h}$. We obtain after some algebra
\begin{equation}
\bm{h}\,{=}\,V\,\Big[\big[\bm{\underline{I}}_{16}\,{-}\,\bm{\underline{P}}_{s}{\otimes}\bm{
\underline{R}}_{t}(\bm{0})\big]\,\underline{\bm{G}}\,{+}\,\bm{\underline{I}}_{16}\Big]^{-1}\bm{1}, \nonumber
\label{Eq:h}
\end{equation}
where $\bm{1}$ is the 16-component column vector of ones. Averaging uniformly over all nonabsorbing
states, i.e.\ over all initial relative coordinates $\bm{r}$ and orientations, we remarkably derive
a closed analytical expression for the MFET $\langle\tau\rangle(p_s,p_t,m_t)$ in relative coordinates
in the two-body active encounter problem:
\begin{equation}
\langle\tau\rangle\,{=}\,\frac{V}{16(V{-}1)}\,\bm{1}^{\top}\underline{\bm{G}}\,
\Big[\big[\bm{\underline{I}}_{16}\,{-}\,\bm{\underline{P}}_{s}{\otimes}\bm{\underline{R}}_{t}(\bm{0})\big]
\,\underline{\bm{G}}\,{+}\,\bm{\underline{I}}_{16}\Big]^{-1}\!\!\!\bm{1}.
\label{Eq:MFET}
\end{equation}

\begin{figure*}[t]
\centering
\includegraphics[width=0.98\textwidth]{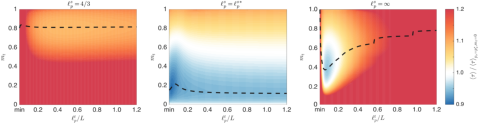}
\caption{Mean first-encounter time (MFET), normalized by the globally optimal value for an immobile
particle $t$ ($m_t{=}0$), as a function of $\ell_p^t$ and $m_t$. The persistence length of particle 
$s$ is set to $\ell_p^s\,{=}\,4{/}3$ (diffusion, left), $\ell_p^s\,{=}\,\ell_p^{s*}$ (middle), and 
$\ell_p^s\,{=}\,\infty$(ballistic motion, right). The color scale ranges from blue to red through 
white, representing MFETs below, equal to, and above the optimal static benchmark, respectively; 
the white level corresponds to $\langle \tau \rangle{/}\langle \tau \rangle_{p_s\,{=}\,p_s^*,m_t
\,{=}\,0}\,{=}\,1$. Results are shown for $V{\simeq}10^4$. The dashed lines indicate the optimal 
mobility $m^{*}_t$, minimizing the MFET for each value of $\ell_p^t$.}
\label{Fig3}
\end{figure*}

\smallskip\smallskip\smallskip
\noindent{\bf Optimization of first-encounter kinetics} 
\smallskip\smallskip\smallskip

\noindent The analytical expression above allows us to systematically explore how persistence and mobility
asymmetries shape the first-encounter kinetics. Throughout this section, we normalize the MFET
by the globally optimal value for an immobile particle $t$ ($m_t\,{=}\,0$), which provides a
natural benchmark for assessing the effect of target mobility. Note that the corresponding
optimal persistence $p_s{=}p_s^*$ for a static target $t$ is system-size dependent \cite{Tejedor12}.
The analytical predictions are validated by extensive Monte Carlo simulations over an ensemble
of $10^6$ realizations for each parameter set. The numerical results agree with the analytical
solution to within a maximum relative error of $0.5\%$ throughout the explored parameter space.

The resulting optimization landscapes in the $(\ell_p^s,\ell_p^t)$ plane for increasing mobility 
$m_t$ are shown in Fig.\,\ref{Fig2}. For an immobile particle $t$, the MFET exhibits a minimum at
$\ell_p^s{/}L{\approx}0.2$, independent of $\ell_p^t$. As $m_t$ increases, a competing configuration 
with high $\ell_p^s$ and low $\ell_p^t$ progressively becomes favorable, leading toward two 
symmetry-related low-high persistence optima as equal mobility is approached. Remarkably, weak 
mobility creates a region in which the MFET falls below the globally optimal static benchmark. 
This sub-benchmark regime subsequently disappears as the mobilities become equal, demonstrating 
that appropriately tuned mobility can enhance encounter efficiency beyond that achievable for 
a static partner.

\begin{figure}[b]
\centering
\includegraphics[width=0.42\textwidth,clip]{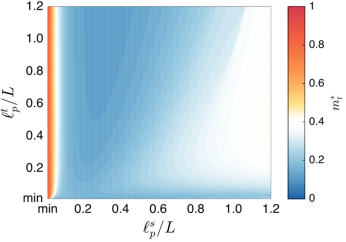}
\caption{Optimal mobility $m^{*}_t$ of the slower particle $t$ in terms of the individual 
persistence values of both particles, $\ell_p^s$ and $\ell_p^t$. The color scale ranges 
from blue for $m^{*}_t\,{=}\,0$ (static particle $t$) to red for $m^{*}_t\,{=}\,1$ (equal 
mobility case).}
\label{Fig4}
\end{figure}

To isolate the mobility-induced effects of particle $t$ on the MFET, we fix $\ell_p^s$ while varying 
the two remaining dynamical variables:\ the persistence $\ell_p^t$ and mobility $m_t$. Examples are 
shown in Fig.\,\ref{Fig3} for the persistence length of particle $s$ being set to $\ell_p^s\,{=}\,4{/}3$ 
(diffusion), $\ell_p^{s*}$, and $\infty$ (ballistic motion). The results reveal that increasing 
mobility can reduce the MFET (even outperforming the globally optimal static benchmark). However, 
pushing $m_t$ further toward equal mobility ($m_t\,{=}\,1$) gradually diminishes this enhancement 
and can degrade encounter efficiency. The optimal mobility $m^{*}_t$, which minimizes the MFET, 
depends nontrivially on the individual persistence values of both particles. Specifically, for 
a diffusive particle $s$, relatively large mobilities optimize the MFET across all persistence 
values of the slower particle $t$. At $\ell_p^s\,{=}\,\ell_p^{s*}$, the minimum MFET occurs at 
smaller values of $m^{*}_t$. Here, the global minimum across all $\ell_p^t$ emerges for weak 
mobility combined with a low $\ell_p^t$ (corresponding to a nearly diffusing particle $t$). As 
the persistence of particle $s$ approaches the ballistic regime, $m^{*}_t$ shifts to intermediate 
values. As can be seen across all panels, the MFET becomes less sensitive to the persistence 
of particle $t$ when it moves very persistently (i.e.\ at large $\ell_p^t$ values). Finally, 
the dependence of $m^{*}_t$ on both persistence parameters is summarized in the phase diagram 
of Fig.\,\ref{Fig4}. It highlights the mobility of particle $t$ as a powerful, nontrivial 
control parameter for optimizing encounter efficiency. It is evident that the choice of the 
individual persistence values of the two particles, $\ell_p^s$ and $\ell_p^t$, strongly 
influences the optimal mobility. 

\begin{figure*}
\centering
\includegraphics[width=0.65\textwidth]{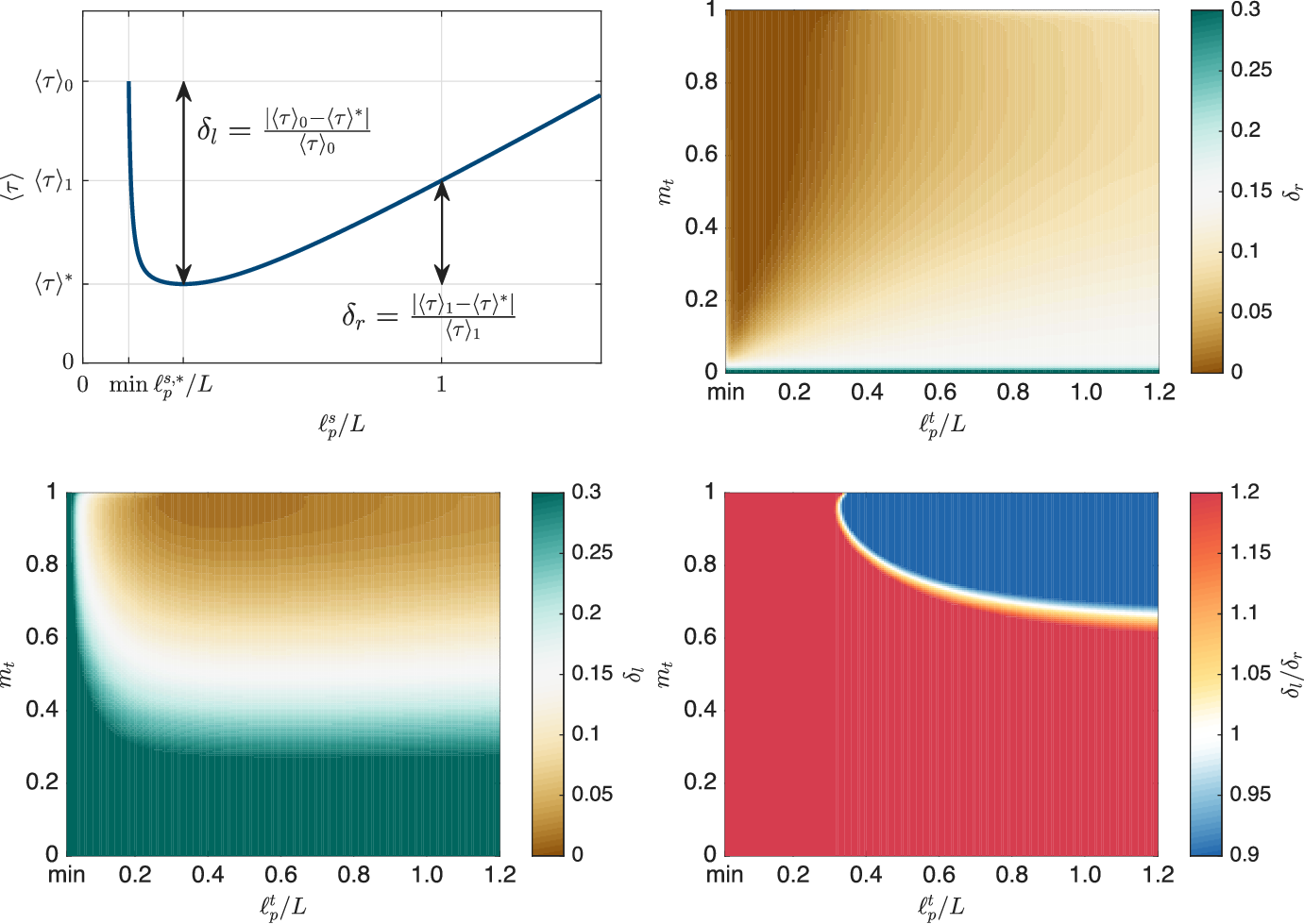}
\caption{(top-left) Definition of the optimization-depth metrics $\delta_l$ and $\delta_r$,
quantifying the fractional gain obtained at the optimal persistence $\ell_p^s\,{=}\,\ell_p^{s*}$
relative to the low and high $\ell_p^s$ limits, respectively. Other panels show
$\delta_r$, $\delta_l$, and the asymmetry ratio $\delta_l{/}\delta_r$ in the
$(\ell_p^t,m_t)$ plane. In the $\delta_l$ and $\delta_r$ maps, brown (green) denotes
a shallow (deep) MFET minimum relative to the corresponding $\ell_p^s$ limit. In the
asymmetry map, red (blue) denotes stronger variation on the low (high) $\ell_p^s$
side of the optimum.}
\label{Fig5}
\end{figure*}

In the above analysis, the reference MFET for comparison is chosen as the one obtained for 
the choice of the optimal persistence $\ell_p^{s*}$ for particle $s$ and an immobile particle 
$t$. However, particle $t$ is generally both mobile and persistent. We therefore ask how robust 
the optimization with respect to $\ell_p^s$ remains when $(m_t,\ell_p^t)$ are varied, and whether
the resulting optimization can be enhanced or suppressed by the mobility of particle $t$.
Figure~\ref{Fig5} quantifies this through the fractional gains $\delta_l$ and $\delta_r$
relative to the low and high $\ell_p^s$ limits, respectively. For $\delta_r$, the optimization
is deep at $m_t\,{=}\,0$ but becomes shallower with increasing mobility or decreasing
persistence, approaching zero for high $m_t$ and low $\ell_p^t$. In contrast, $\delta_l$
crosses from deep to shallow with increasing $m_t$ and is largely insensitive to $\ell_p^t$
except at very small $\ell_p^t$. The ratio $\delta_l{/}\delta_r$ shows stronger variation
on the low $\ell_p^s$ side over most of the parameter space, with the asymmetry reversing
only for simultaneously high $m_t$ and $\ell_p^t$. These results confirm that the mobility-induced
enhancement regimes in Fig.\,\ref{Fig3} persist beyond the static-target reference and
remain robust to variations in $\ell_p^s$.

\smallskip\smallskip\smallskip
\noindent{\bf Equal mobility regime} 
\smallskip\smallskip\smallskip

\noindent The general solution admits further analytical simplifications in several physically relevant
limiting cases. Particularly, in the major sub-class of equal mobility $m_t\,{=}\,1$, the
relative dynamics acquires additional symmetry. In this regime, $\bm{\underline{R}}_{t}(\bm{0})$
and $\bm{h}$ reduce to $\bm{\underline{R}}_{t}(\bm{0})\,{=}\,\bm{\underline{P}}_{t}$ and $\bm{h}
\,{=}\,V\,\Big[\big[\bm{\underline{I}}_{16}\,{-}\,\bm{\underline{P}}_{s}{\otimes}\bm{\underline{P}}_{t}
\big]\,\underline{\bm{G}}\,{+}\,\bm{\underline{I}}_{16}\Big]^{-1}\bm{1}$, leading to a simpler
expression for $\langle\tau\rangle$. Of particular interest is the two-diffusing-particle
limit \cite{Moreau04}, $p_s\,{=}\,p_t\,{=}\,0$, for which we find
\begin{equation}
\langle\tau\rangle_{_\text{diffusive}}\,{=}\,\displaystyle\frac12\frac{V}{V{-}1}\,\sum\limits_{\bm{k}\,
{\neq}\,\bm{0}}\Big(\frac{1}{1{-}\gamma(\bm{k})}\,{+}\,\frac{1}{1{+}\gamma(\bm{k})}\Big), \nonumber
\label{Eq:MFET-BothDiffusive}
\end{equation}
where $\gamma(\bm{k})\,{=}\,\frac12(\cos k_x{+}\cos k_y)$. For comparison, a static-diffusive
pair, $m_t\,{=}\,p_t\,{=}\,p_s\,{=}\,0$, on a discrete lattice with periodic setting yields
$\langle\tau\rangle_{_\text{static}}\,{=}\,\displaystyle\frac{V}{V{-}1}\,\sum_{\bm{k}\,{\neq}
\,\bm{0}}\frac{1}{1{-}\gamma(\bm{k})}$. The two MFETs become asymptotically equivalent in the
large system limit $L\,{\rightarrow}\,\infty$, $\lim\limits_{L{\rightarrow}\infty} \langle\tau
\rangle_{_\text{diffusive}}{/}\langle\tau\rangle_{_\text{static}}\,{=}\,1$. This equivalence
need not necessarily persist in the presence of interparticle interactions, geometric
inhomogeneity, obstacles, or nonperiodic boundaries.

\begin{figure*}
\centering
\includegraphics[width=0.98\textwidth]{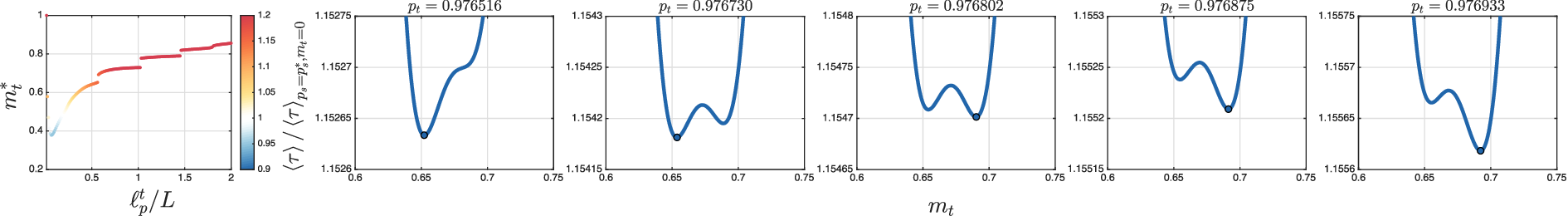}
\caption{Optimal target mobility $m_t^{\ast}$ vs target persistence length
$\ell_p^t$ for fixed $p_s\,{=}\,1$. (left) Discontinuous jumps in $m_t^{\ast}$ arise from
transitions between competing mobility optima. Other panels:\ Cuts through the normalized
MFET landscape at selected $\ell_p^t$, showing competing local minima and their spontaneous
exchange as the global minimum. Circles indicate the global minima.}
\label{Fig6}
\end{figure*}

\smallskip\smallskip\smallskip
\noindent{\bf Discontinuous optimal mobility} 
\smallskip\smallskip\smallskip

\noindent A remarkable feature emerges in the ballistic limit of particle $s$. Its deterministic motion
leaves $m_t$ and $\ell_p^t$ of particle $t$ as the remaining degrees of freedom. In this case,
directional changes of particle $t$ can lead to interesting encounter scenarios. For instance,
by fixing $\ell_p^t$, the optimal mobility ${m}_t^*$ can be determined by $\partial\langle\tau
\rangle{/}\partial m_t|_{_{m_t{=}{m}_t^*}}\,{=}\,0$. As shown in Fig.\,\ref{Fig6}, ${m}_t^*$
varies discontinuously with $\ell_p^t$, revealing abrupt transitions between competing encounter
strategies. Introducing $\bm{\underline{C}}\,{=}\,m_{t}\left[\bm{\underline{I}}_{4}{-}\bm{
\underline{P}}_{t}\right]\underline{\bm{G}}\,{+}\,\bm{\underline{I}}_{4}$ and its derivative
$\bm{\underline{C}}'$ with respect to $m_t$, the minimum MFET at ${m}_t^*$ fulfills $\sum\limits_{
\otf}\bm{1}^{\top}(\underline{\bm{G}}'\,{-}\,\underline{\bm{G}}\,\bm{\underline{C}}^{-1}\,
\bm{\underline{C}}')\bm{h}\,{=}\,0$. The example cuts of MFET in Fig.\,\ref{Fig6} show
that the jumps of ${m}_t^*$ arise from the exchange of competing local minima:\ as $\ell_p^t$
is varied, a secondary minimum becomes globally optimal and subsequently loses its advantage.
Thus, the discontinuities in the optimal mobility originate from a nontrivial restructuring
of the MFET landscape rather than from a singularity in the underlying dynamics.

\smallskip\smallskip
\noindent{\bf Conclusions} 
\smallskip\smallskip

\noindent We have developed an exact analytical framework for the general class of first-encounter kinetics 
problems of independently moving active particles, overcoming the difficulty that their individual 
persistence precludes a conventional reduction to a single effective walker. Our results reveal 
that mobility is not merely an additional source of relative diffusion, but a control parameter 
that reshapes the optimization landscape and, when suitably tuned, can enhance encounter efficiency 
beyond the globally optimal static benchmark. Optimal strategies are generally asymmetric in 
persistence, while competing optima and discontinuous transitions emerge in distinct dynamical 
regimes. The excellent agreement with Monte Carlo simulations demonstrates the predictive power 
of the formalism. More broadly, this framework provides a basis for studying first-encounter 
kinetics among multiple active agents in confined environments, with relevance to biological, 
ecological, and synthetic systems, and opens avenues toward heterogeneous motility, interparticle 
interactions, and complex boundaries.

\bibliography{Refs-ActiveEncounter}

\end{document}